\documentclass{sf2a-conf2013}
\usepackage{graphicx}
\usepackage{hyperref}
\usepackage[]{natbib}  
\usepackage{epstopdf}
\newcommand{\teff}{T_\mathrm{eff}}

\def\BibTeX{{\rm B\kern-.05em{\sc i\kern-.025em b}\kern-.08em
    T\kern-.1667em\lower.7ex\hbox{E}\kern-.125emX}}
\bibpunct{(}{)}{;}{a}{}{,}  %%%%%%%%%%%%%  A&A bibliography style
\begin{document}

\TitreGlobal{SF2A 2013}

%%-----------------------------------------------------------------
%%      the top matter
%%

\title{High resolution spectroscopy of M subdwarfs}

\runningtitle{High resolution spectroscopy of M subdwarf}

\author{A. S. Rajpurohit}\address{Institut UTINAM CNRS 6213, Observatoire des Sciences de l'Univers THETA Franche-Comt\'{e}-Bourgogne, Universit\'e de Franche Comt\'{e}, 
Observatoire de Besan\c{c}on, BP 1615, 25010 Besan\c{c}on Cedex, France}

\author{C. Reyl\'e$^1$}
\author{M. Schultheis}\address{Universit\'{e} de Nice Sophia-Antipolis, CNRS, Observatoire de C\^{o}te d'Azur, Laboratoire Cassiop\'{e}e, 06304 Nice Cedex 4, France}
\author{F. Allard}\address{Centre de Recherche Astrophysique de Lyon, UMR 5574: CNRS, Universit\'{e} de Lyon, \'{E}cole Normale Sup\'{e}rieure de Lyon, 46 all\'{e}e d'Italie,
69364 Lyon Cedex 07, France}
\setcounter{page}{237}

%%-----------------------------------------------------------------

\maketitle

%%-----------------------------------------------------------------
%%        The abstract
%% 
%%  Warning!  within the abstract:
%%  - do not use macros. 
%%  - do not use commands like: \cite, \citet, \citep ... etc.

\begin{abstract}
M subdwarfs are metal poor and cool stars. They are important probes of the old galactic populations. However, they remain elusive due to their low luminosity. Observational and modeling efforts are required to fully understand the physics and to investigate the effect of metallicity in their cool atmospheres. We perform a detail study of a sample of subdwarfs to determine their stellar parameters and constrain the atmosphere models.
We present UVES/VLT high resolution spectra of 21 M subdwarfs. Our atlas covers the optical region from 6400$\AA$ up to the near infrared at 10000$\AA$. We show spectral details of cool atmospheres at very high resolution (R$\sim$ 40 000) and compare with synthetic spectra computed from the recent BT-Settl atmosphere models. Our comparison shows that molecular features (TiO, VO, CaH), and atomic features (Fe, Ti, Na, K) are well fitted by current models. We produce an effective temperature versus spectral type relation all over the subdwarf spectral sequence.

\end{abstract}

%% Insert the keywords (to appear in the ADS indexing)
%% Keywords must be separated by a comma

%%-----------------------------------------------------------------

\section{Introduction}
%%---------------------
Due to their lower metallicity and intrinsic faintness, M subdwarfs lie below their solar-metallicity counterpart in the Hertzsprung-Russel diagram. Subdwarfs are typically very old (10 Gyr or more) and belonging to the old Galactic populations: old disc, thick disc and spheroid, as shown by their spectroscopic features, kinematics properties and ages \citep{Digby2003,Lepine2003a, Burgasser2003}.  Detailed studies of their complex spectral energy distributions give new insights on the role of metallicity in the opacity structure, chemistry and evolution of cool atmospheres, and fundamental issues on spectral classification and temperature versus luminosity scales.  \cite{Gizis1997} proposed a first classification of M subdwarfs (sdM) and extreme subdwarfs (esdM) based on TiO and CaH band strenghts in low resolution optical spectra. \cite{Lepine2007} has recently revised the adopted classification, and proposed a new classification for the most metal poor, the ultra subdwarfs (usdM).  \cite{Jao2008} compared model grids with the optical spectra to characterize the spectral energy distribution of subdwarfs by three parameters: temperature, gravity, and metallicity, and thus gave an alternative classification scheme of subdwarfs.  Metallicities have been recently obtained using high resolution spectra by measuring equivalent width of atomic lines in regions not dominated by molecular bands \citep{Bean2006,Woolf2005,Woolf2009}\\

Rapid progresses in the investigation of cool atmospheres are expected thanks to the advent of 8 m class telescopes that allow high resolution spectroscopy of these faint targets. In high resolution spectra, access to weak lines allows us to determinate the metallicity disentangled from the other main parameters, gravity and temperature. The pressure changes affect equally all atmospheric parameters and therefore the various absorption bands. The determination of gravity from the pressure broadened wings can be expected to be much more accurate than comparing colour ratios from photometry and/or low resolution spectra.  Thus it is necessary to achieve a very good fit in all important absorbers in order to determine atmospheric properties, because the chemical complexity of these atmospheres reacts sensibly to the major opacities. Descriptions of these stars therefore need validations by comparing with high resolution spectroscopic observations. \\

In this paper we present the first high resolution optical atlas of stars covering the whole subdwarf sequence. It consists of 21 sdM, esdM, and usdM observed with UVES at VLT. Using the most recent \texttt{PHOENIX} BT-Settl stellar model atmosphere we have performed a detailed comparison with our observed spectra using a $\chi^2$ minimization technique. In this study we confront the models to the high resolution spectra of subdwarfs and we assign effective temperatures to the M dwarfs. We derive metallicities based on the best fit of synthetic spectra to the observed spectra and perform a detailed comparison of lines profile of individual elements such as Fe I, Ca II, Ti I, Na.

\section{Observation and data reduction}
%%---------------------------------

The observation were carried out in visitor mode during April and September 2011 with the optical spectrometer UVES \citep{Dekker2000} on the Very Large Telescope (VLT) at the European Southern Observatory (ESO) in Paranal, Chile. In total 21 targets were observed during period 87. UVES was operated in dichroic mode using red arm with non-standard setting centered at 830 nm. This setting covers the wavelength range 6400 $\AA$-9000 $\AA$, which contains various atomic lines like Fe I, Ti I, K II, Na I and Ca II and is very useful for the spectral synthesis analysis. The spectra were taken with a slit width of 1.0" yielding a nominal resolving power of R = $\lambda$/{$\Delta$$\lambda$} = 40 000. The signal to noise ratio varies over the wavelength region according to the object's spectral energy distribution and detector efficiency. It reaches 30 in most of the spectral region (6400 $\AA$-9000 $\AA$) so that the dense molecular and atomic absorption features are well discernible from noise. Data were reduced using the ESO based software called REFLEX for UVES data which uses standard ESO pipeline modules.

\section{Model Atmosphere}
In this study, we use the most recent BT-Settl models partially published in a review by \cite{Allard2012}. These model atmospheres are computed with the \texttt{PHOENIX} multi-purpose atmosphere code version 15.5 \citep{Hauschildt1997,Allard2001} solving the radiative transfer in 1D spherical symmetry, with the classical assumptions: hydrostatic equilibrium, convection using the mixing length theory, chemical equilibrium, and a sampling treatment of the opacities. The models use a mixing length as derived by the radiation hydrodynamic) simulations of \cite{Ludwig2002, Ludwig2006} and \cite{Freytag2012} and a radius as determined by the \cite{Baraffe1998} interior models as a function of the atmospheric parameters ($\teff$, $\mathrm{log}\,g$, [Fe/H]).\\

The BT-Settl grid extends from $\teff$ = 300 to 7000 K at a step of 100 K, $\mathrm{log}\,g$ = 2.5 to 5.5 at a step of 0.5 dex and [Fe/H]= -2.5 to 0.0 at a step of 0.5 dex accounting for alpha element enrichment. We interpolated the grid at every 0.1 dex in $\mathrm{log}\,g$ and metallicity. For more detail of BT-Sett model atmosphere see \cite{Allard2012, Rajpurohit2012a}. The synthetic colours and spectra are distributed with a spectral resolution of around R=100000 via the \texttt{PHOENIX} web simulator\footnote{http://phoenix.ens-lyon.fr/simulator}. \\

%Fig. \ref{Fig:3} shows BT-Settl synthetic spectra varying $\teff$ and [M/H].  It shows that the oxide bands that dominate in M dwarfs spectra are weaker in the subdwarfs where the hydrides bands dominate\citep{Jao2008}. They have complex and extensive band structures leaving no window for the true continuum and creating pseudo continuum that only let through the strongest, often resonance atomic lines \citep{Allard1990, Allard1995}. However because of the lower metallicity of subdwarfs, the TiO bands are less strong, and the pseudo-continuum is brighter. This increases the contrast to the other opacities such as hydride bands and atomic lines which feel the higher pressures of the deeper layers where they emerge from. We see therefore these molecular bands with more details and under more extreme gas pressure conditions than for M dwarfs.
%
%\begin{figure*}[ht!]
%\centering
%\includegraphics[width=13.5cm,height=10.0cm]{CaH.ps}
%\caption{ BT-Settl synthetic spectra from 4000 K to 3000 K. The black, red, and blue lines represent [M/H] = 0.0, -1.0, and -2.0, respectively. %The stronger CaH bands at a given temperature could be caused by lower metallicity. This should not be in the caption.
%}
%\label{Fig:3}
%\end{figure*}

\section{Comparison with model atmospheres}

\begin{figure*}[ht!]
\centering
\includegraphics[width=15.5cm]{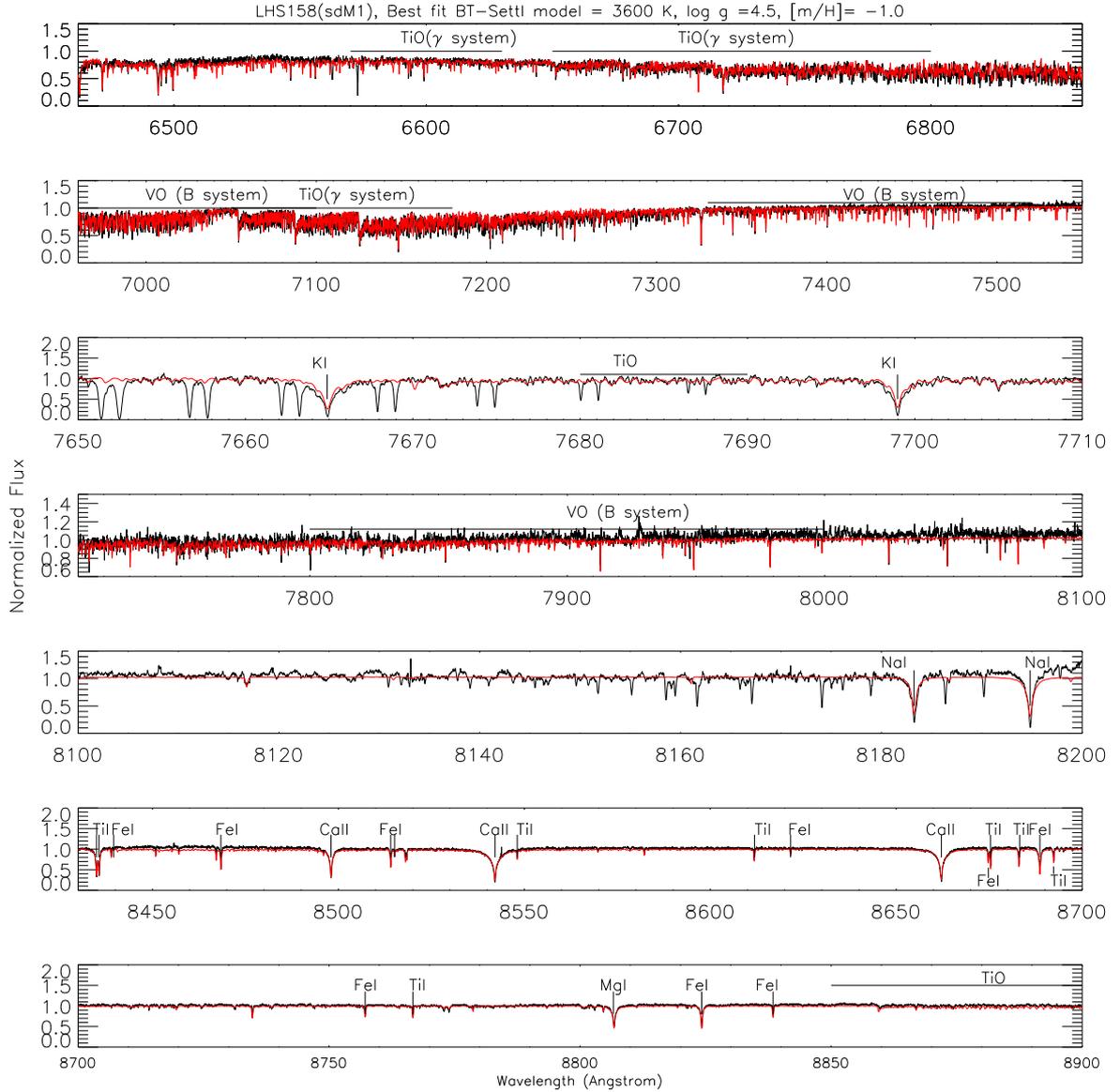}
\caption{UVES spectra of the sdM1 star LHS~158 (black) compared to the best fit BT-Settl synthetic spectra (red).}
\label{Fig:4}
\end{figure*}

We perform a comparison between observed and synthetic spectra computed from the BT Settl model to derive the physical parameters of our sample.
Furthermore, the comparison with observed spectra is very crucial to reveal the inaccuracy or incompleteness of the opacities used in the model.  Fig. \ref{Fig:4} shows the comparison of the best fit model for a sdM1 star.

The analysis using synthetic spectra requires the specification of several inputs parameters: effective temperature, surface gravity and the metallicity compared to Sun. As opposed to previous studies in which the best fit was found by trial and error, in this paper we derive the effective temperature by performing a $\chi^2$ minimization technique.  Our procedure is a two-step process. We first convolve the synthetic spectrum with a Gaussian kernel at the observed resolution and then rebin the outcome with the observation. For each of the observed spectra we compare these spectra with the grid of synthetic spectra in the wavelength range 6400$\AA$ to  9000$\AA$ using $\chi^2$ minimization technique.  We have excluded the spectral region between 6860 to 6960 $\AA$, 7550 to 7650 $\AA$, and 8200 to 8430 $\AA$, due to presence of atmospheric absorption. In the first step, we keep all the stellar parameter ($\teff$, [Fe/H], $\mathrm{log}\,g$) free.\\

In a second step, a  $\chi^2$ map is obtained for each of the observed spectra as function of temperature and metallicity. The error bar are derived from standard deviation by taking 5 $\%$ from the minimum $\chi^2$ value. The acceptable parameters were finally inspected by comparing it with the observed spectra.  We found generally good agreement with BT-settl model and conclude that model fitting procedure can be used to estimate the $\teff$ with an accuracy better than $\sim$100\, K.  

\begin{figure}[ht!]
\centering
\includegraphics[width=8cm]{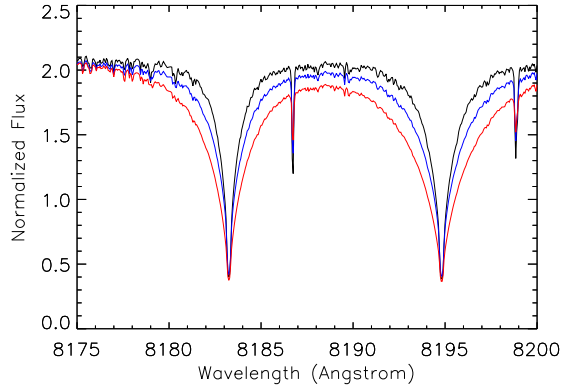}
\caption{BT-Settl synthetic spectra with $\teff$ of 3500 K and varying $\mathrm{log}\,g$ =4.5 (black), 5.0 (blue), 5.5 (red). The effect of gravity and pressure broadening on the sodium doublet is clearly visible.}
\label{Fig:NaI}
\end{figure}

We tested the sensitivity of the TiO lines to surface gravity and did not find any significant effect in the parameter space and wavelength range relevant to this study.  The surface gravity can be determined by analyzing the width of gravity sensitive atomic lines such as K I and Na I doublets (see Fig. \ref{Fig:NaI}) as well as relative strength of metal hydride bands such as CaH. The overall line strength (central depth and equivalent width) increases with gravity as the decreasing ionization ratio due to higher electron pressure leaves more neutral alkali lines in the deeper atmosphere. The width of the damping wings in addition increases due to the stronger pressure broadening, mainly by H$_2$, He and H I collision. 

\begin{figure*}[ht!]
\centering
\includegraphics[trim=0cm 3.5cm 0cm -0.5cm, width=13.5cm,height=10.0cm]{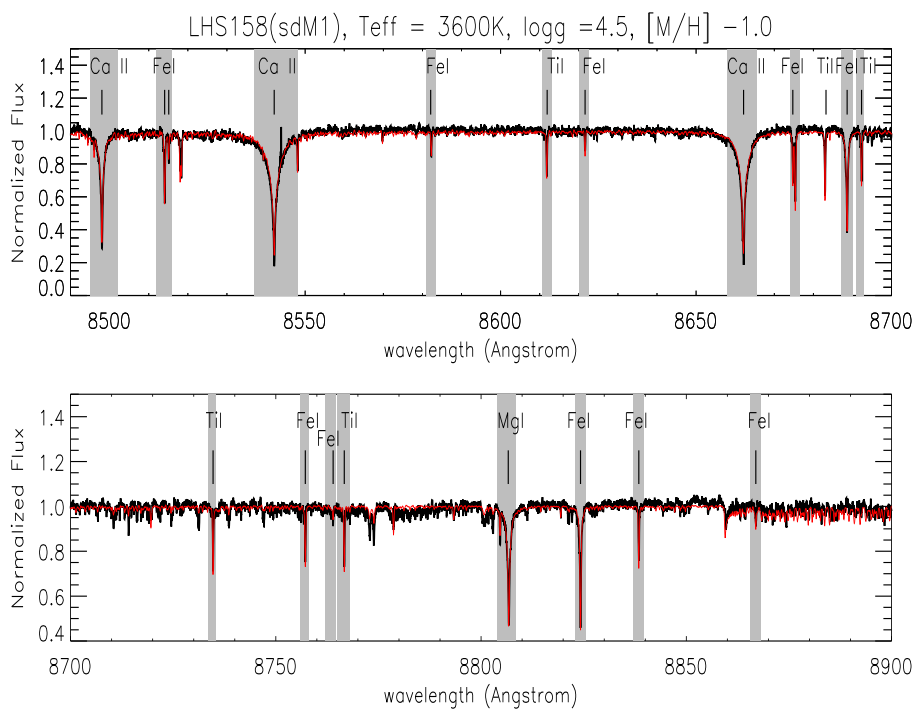}
\caption{UVES spectrum of the sdM1 star LHS 158 (black) and the best fit BT Settl synthetic spectrum (red). The spectral region used to determine metallicity is shown only. The atomic features used are highlighted.}
\label{Fig:5}
\end{figure*}

For the metallicity determination we fit the synthetic spectra using the same procedure explained above but on restricted regions where molecular absorptions are less strong and atomic lines appear clearly. Our first criterion is that the lines must have fractionally small amount of TiO line blending. This necessitate the use of fairly strong lines. Most of the lines we select occupy a short spectral interval from 8440 $\AA$ to  8900 $\AA$ (see fig. \ref{Fig:5}). In our analysis we also include the new lines of elemental species such as Ca II triplet, Ti lines, and TiO bandhead around 7088 $\AA$. The best fit parameters ($\teff$, $\mathrm{log}\,g$, [Fe/H]) are given in Table 3. 
\begin{table*}
\centering
\caption{Stellar Parameters of the observed targets.}
\begin{tabular}{ccccc}
\hline
Target& Spectral Type &   $\teff$ (K) &	  $\mathrm{log}\,g$ &	  [Fe/H]\\
\hline

LHS 72&				sdK4	&	3900	$\pm$23	&4.5$\pm$0.13&	-1.4$\pm$	0.27 \\
LHS 73&				sdK7	&	3800	$\pm$39	&4.5$\pm$0.13&	-1.4$\pm$0.19\\
G 18-37&			esdK7&	3800	$\pm$78	&4.5$\pm$0.15&	-1.3$\pm$0.44\\%SDSS 221455
APMPM J2126-4454&		sdM0&	3700	$\pm$49	&4.5	$\pm$0.19&	-1.3$\pm$0.23\\
LHS 300&				sdM0&	3800	$\pm$39	&4.5$\pm$0.17&	-1.4$\pm$0.24\\
LHS 401&				sdM0.5&	3800	$\pm$26	&4.5$\pm$0.17&	-1.4$\pm$0.28\\
LHS 158&				sdM1&	3600	$\pm$48	&4.5$\pm$0.17&	-1.0$\pm$0.3\\
LHS 320&				sdM2&	3600	$\pm$59	&4.6$\pm$0.23&	-0.6$\pm$0.31\\
LHS 406&				sdM2&	3600	$\pm$40	&4.7$\pm$0.24&	-0.6$\pm$0.24\\
LHS 161&				esdM2&	3700	$\pm$77	&4.8$\pm$0.30&	-1.2$\pm$0.36\\ %sdK6.0
LP 771-87&		usdM2&	3600	$\pm$95	&4.8$\pm$0.31&	-1.4$\pm$0.52\\%2MASS 030734
LHS 541&			sdM3&	3500	$\pm$76	&5.1$\pm$0.31&	-1.0$\pm$0.39\\
LHS 272&				sdM3&	3500	$\pm$66	&5.2$\pm$0.30&	-0.7$\pm$0.37\\
LP 707-15&			esdM3&	3500	$\pm$68	&5.5$\pm$0.29&	-0.5$\pm$0.36\\%SDSS 010954
LSR J1755+1648&			sdM3.5&	3400	$\pm$52	&5.1$\pm$0.31&	-0.5$\pm$0.33	\\			
LHS 375&				sdM3.5&	3500	$\pm$79	&5.5$\pm$0.32&	-1.1$\pm$0.31\\
LHS 1032&			esdM4&	3300	$\pm$63	&4.5$\pm$0.32&	-1.7$\pm$0.25\\
SSSPM J0500-5406&		esdM6.5&	3200$\pm$51	&5.4$\pm$0.31&	-1.6$\pm$0.16\\
LHS 377&				sdM7&	3100	$\pm$32	&5.3$\pm$0.25&	-1.0$\pm$0.16\\
APMPM 0559-2903&		esdM7&	3200	$\pm$68	&5.4$\pm$0.34&	-1.7$\pm$0.25\\
SSPM J1013-1356&			sdM9.5&	3000	$\pm$0		&5.5$\pm$0.05&	-1.1$\pm$0.16\\

\hline
\end{tabular}
\end{table*}

\section{Discussion}
\label{ccl}

We use the most recent BT-Settl model atmospheres, with revised solar abundances, to determine the scaled solar abundances of the subdwarfs.  
We compare the synthetic spectra produced the model atmospheres and derive their fundamental stellar parameters.  The accuracy of the atmospheric models involved in the metallicity determination can be inferred by looking the fit to the individual atomic and molecular lines. Working with these high resolution spectra allowed us to disentangle the atmospheric parameters (effective temperature, gravity, metallicity), which is not possible when using braodband photometry. With this study we were able to constrain for the first time the BT-Settl models at sub solar metallicities. 

\begin{figure}[ht!]
\centering
\includegraphics[width=9.0cm,height=6.0cm]{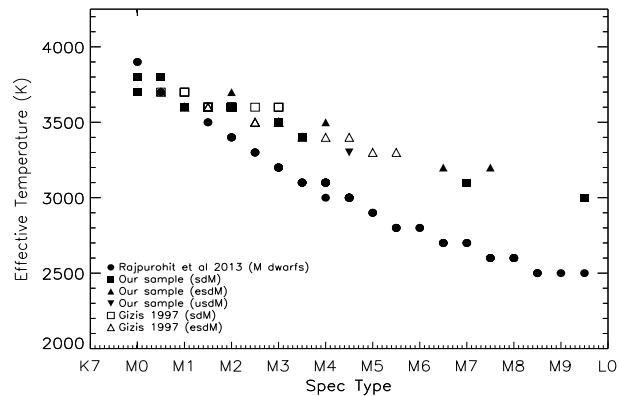}

\caption{Effective temperature of subdwarfs versus spectral type relation from our sample (filled symbols) compared to the one from \cite{Gizis1997} (open symbols) and to the M dwarfs $\teff$ scale from \cite{Rajpurohit2013} (filled circles).}
\label{Fig:teff}
\end{figure}

The effective temperature versus spectral type relation is shown in Fig. \ref{Fig:teff}. The relation determined using UVES sample is compared to the $\teff$ scale of M dwarfs determined by \cite{Rajpurohit2013}. $\teff$ of subdwarfs dwarfs is 200-300 K higher than $\teff$ of M dwarfs for the same spectra type except for hot temperature.  This is expected since the TiO bands are depleted with decreasing metallicity, and as a result the pseudo-continuum is brighter and the flux is emitted from the hot deeper layer. A comparison to the earlier work from \cite{Gizis1997} is also shown. \cite{Gizis1997} determined the temperature by comparing the low resolution optical spectra of a sample of sdM and esdM with the NextGen model atmosphere grid by \cite{Allard1997}. 
The $\teff$ scales are in agreement within 100 K. This difference is due to the incompleteness of the TiO and water vapor line lists used in the NextGen model atmospheres compared to the new BT-Settl models. Furthermore this work allows us to extend the relation to the coolest M subdwarfs.

%% The following lines are required when using BibTEX (strongly encouraged!):
\bibliographystyle{aa}  % A&A bibliography style file (aa.bst)
\bibliography{ref} % your references in file: Yourfile.bib

\end{document}